\documentclass[conference]{sig-alternate-10pt}
\usepackage{color, times}
\usepackage{subfigure, epsf, graphics, graphicx, url}
\usepackage{algorithm,algorithmic,amsmath}
\usepackage{lscape}

\long\def\symbolfootnote[#1]#2{\begingroup%
\def\thefootnote{\fnsymbol{footnote}}\footnotetext[#1]{#2}\endgroup}

\begin{document}
\title{GDB: Group Distance Bounding Protocols}

\author{Srdjan Capkun\\
ETH Zurich\\
capkuns@inf.ethz.ch\\
\and
Karim El Defrawy \\
(corresponding author)\\
UC Irvine\\
keldefra@ics.uci.edu\\
\and
Gene Tsudik\\
UC Irvine\\
gts@ics.uci.edu\\
}

\maketitle

\begin{abstract}
Secure distance bounding (DB) protocols allow one entity, the verifier, to 
securely obtain an upper-bound on the distance to another entity, the prover. 
Thus far, DB was considered mostly in the context of a single 
prover and a single verifier. There has been no substantial prior work on 
secure DB in group settings, where a set of provers 
interact with a set of verifiers. The need for group distance bounding 
(GDB) is motivated by many practical scenarios, including: group device 
pairing, location-based access control and secure distributed localization. 
GDB is also useful in mission-critical networks and automotive computer
 systems. This paper addresses, for the first time, GDB protocols by utilizing 
the new \emph{passive DB} primitive and the novel 
\emph{mutual multi-party GDB} protocol. We show how 
they can be used to construct secure and efficient GDB 
protocols for various settings. We analyze security and performance of our protocols and 
compare them with existing DB techniques when applied to group settings.

\end{abstract}


\section{Introduction}
\label{sec:intro}
Wireless networks -- especially, sensor and mobile ad-hoc networks, have 
become increasingly popular. Enabled by pervasive availability of 
location information, new wireless scenarios have emerged where accurate proximity 
information is essential to both applications and basic networking functions. 
Such scenarios require secure, reliable and efficient verification of 
distances between nodes, in addition to node authentication. 
Distance Bounding (DB) can address such scenarios by allowing one 
entity (verifier) to obtain an upper-bound on the distance to another 
entity (prover) and, optionally, authenticate the latter. DB was introduced by 
Brands and Chaum \cite{chaum-db} as a means of preventing the so-called
``mafia fraud" attacks on bank ATMs\footnote{A ``mafia fraud'' attack occurs
when the attacker identifies itself to the verifier using the identity of a 
prover, without the latter being aware (i.e., man-in-the-middle attack).}. 
In Brands and Chaum's DB approach, a user's smart-card (verifier) 
checks its proximity to the ATM (prover). DB has been recently implemented 
\cite{kasper-db-realization-ss2010} using commercial off-the-shelf 
electronics (resulting in 15cm accuracy). It was also suggested and implemented as a 
means of securely determining node locations in wireless networks 
\cite{Kuhn10uwbdb,capkun-infocom05,db-wisec09,sec-db-loc-preneel}.  

In most prior work, DB was considered in the context of a 
single prover and a single verifier. Group Distance Bounding 
(GDB) is the natural extension of the DB concept to 
group settings with multiple provers and verifiers. Multiple verifiers 
provide several advantages including: higher attack resilience and improved 
availability (by avoiding a single point of compromise or failure), in 
addition to facilitating localization using multilateration. The common 
goal in applications that require GDB is: 
\textit{several devices must securely measure distances between themselves 
or should only operate in the vicinity of each other.} 

GDB is motivated by the following emerging wireless applications:
group device pairing -- a procedure for 
setting up an initial secure channel among a group of previously unfamiliar
wireless devices. There are several scenarios where this is required, e.g., when 
an ephemeral ad-hoc group of users meet. Each user has a personal wireless device that must establish 
a secure channel with devices of other users. One concrete example using cell-phones 
is described in \cite{gangs}. Another scenario is that of a single user with multiple devices, e.g, 
in a home area network \cite{HAN}. A secure mechanism is required to ensure that the group of communicating devices is clustered 
within a particular area, i.e., each device is within a certain 
distance from every other device. The mutual multi-party GDB 
protocol (Section \ref{subsec:mutualmpDB}) achieves this.

Another application that can benefit from GDB is automotive computer systems. 
Recent research \cite{oakland-automotive-sec-2010} 
pointed out vulnerability of such systems to attacks through 
wireless interfaces (demonstrated in \cite{francillion10} using relay attacks). As more components of such 
systems communicate wirelessly, it becomes critical to ensure 
that the origin of communication is from within the car to prevent relay attacks. 
Ensuring that such components only communicate with each 
other prevents attacks through unauthorized or outside malicious 
components. The mutual multi-party 
GDB protocol (Section \ref{subsec:mutualmpDB}) achieves this.

GDB is also useful in critical, e.g., military, MANETs where a key operational 
requirement is to track locations of, and authenticate, friendly nodes 
\cite{mil-manet-req}. Critical MANETs generally operate without any 
infrastructure and in hostile environments where node compromise is quite realistic. 
GDB can be used to implement location based-access control and location-based group 
key management in critical MANETs. Both mutual multi-party GDB (Section \ref{subsec:mutualmpDB}) 
and passive DB (Section \ref{subsec:passive-db}) can be used in such settings.

In this paper, we show that a straightforward extension of previous 
single prover single verifier DB to GDB is inefficient and
 insecure if used for localization without synchronization 
between verifiers (which was also pointed out in \cite{db-wisec09}). We explore 
and propose more efficient and secure GDB approaches. We make the 
following contributions:\\
\textit{- Definition of Group Distance Bounding (GDB)}\\
\textit{- New primitives: Passive DB and Mutual Multi-Party GDB}\\
\textit{- A set of secure and efficient GDB protocols}\\
\textit{- Security and performance analysis of proposed protocols}

The rest of the paper is organized as follows: we overview traditional DB protocols, 
formulate the GDB problem and state our system and 
adversary models in Section \ref{sec:bkgrnd-prblem-stmnt}. We present details and 
security analysis of our GDB building block primitives in Section \ref{sec:gdb-building-blocks}. We then show how to use these 
building blocks to construct GDB protocols in both one-way and mutual GDB settings in Section 
\ref{sec:gdb-protocols}. We analyze performance and security of 
GDB protocols in Section \ref{sec:perf-sec-analysis}. We discuss related work in Section 
\ref{sec:related-work} and conclude with open issues and future work in Section \ref{sec:disc-future-work}.

\section{Background and Problem Statement}
\label{sec:bkgrnd-prblem-stmnt}
We begin with an overview of DB protocols, followed by the problem statement and system model.

\subsection{Overview of Distance Bounding (DB)}
\label{subsec:db-overview}
\begin{figure}[t]
\begin{center}
\fbox{\centering
\includegraphics[width=0.42\textwidth, height=0.21\textheight]{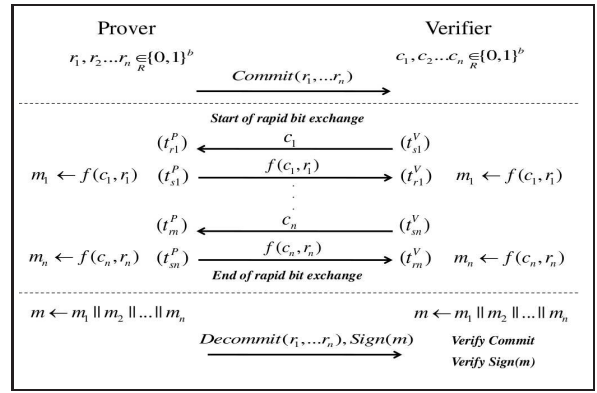}
}
\end{center}
\caption{Basic DB Operation.}
\label{fig:basic-db}
\end{figure}

Figure \ref{fig:basic-db} shows the generic DB protocol operation. The
core of any \textit{one-way DB} protocol is the distance measurement phase, 
whereby the verifier measures round-trip time between sending its challenge 
and receiving the reply from the prover. 
Verifier's challenges are unpredictable to the prover and replies are computed as 
a function of these challenges. Thus, the prover cannot reply to the verifier 
before receiving its challenges. The prover, therefore, cannot pretend to be 
closer to the verifier than it really is (only further). First, the verifier 
and the prover each generate $n$ $b$-bit nonces $c_i$ and $r_i$ ($1\le~i\le~n$), 
respectively. In the Brands-Chaum DB protocol \cite{chaum-db}, the prover also commits 
to its nonces (using any secure commitment scheme). The verifier sends all $c_i$ to the prover, one at a time. Once each 
$c_i$ is received, the prover computes, and responds with a function of its own nonce and 
that of the verifier, $f(c_i,r_i)$. The verifier checks the reply and measures the 
elapsed time between each challenge and response. The process is repeated $n$ times and the 
protocol completes successfully only if \textit{all} $n$ rounds succeed and all 
responses correspond to prover's committed value. The processing time on the prover's side 
$\alpha=t_{s}^P-t_{r}^P$ must be negligible (compared to the time of flight); otherwise, 
a computationally powerful prover could claim a false bound. This time might be tolerably small,
depending on the underlying technology, the distance measured and the 
required security guarantees (less than $1nsec$ processing 
time yields $0.15m$ accuracy \cite{kasper-db-realization-ss2010}).

Security of DB protocols relies on two assumptions: (1) challenges are random, and
unpredictable to the prover before being sent by the verifier, and (2) challenges traverse
the distance between the prover and the verifier at maximum possible speed. i.e., the 
speed of electromagnetic waves. After executing a DB protocol, the verifier knows 
that distance to the prover is at most $\frac{t_r^V - t_s^V-\alpha}{2}\cdot c$,
where $\alpha$ is the processing time of prover (ideally,  negligible) and $c$ the speed of 
light \cite{chaum-db}. DB protocols typically require $(2n + {\cal C})$ messages, 
where ${\cal C}$ is the number of 
messages exchanged in the pre- and post-processing protocol phases. 
Typically, ${\cal C} << n$ and thus can be ignored.

In some cases (e.g., distributed localization), there is a need for mutual DB between 
two parties: $P_1$ and $P_2$. This can be achieved by modifying the one-way DB protocol 
such that 
each response from $P_2$ to a challenge by $P_1$ also includes a challenge from 
$P_2$ to $P_1$. This requires $2n + 2{\cal C} + 1$ messages instead of $2(2n + {\cal C})$ 
for mutual DB and is shown in \cite{sector}. Both parties generate and commit to two random bit 
strings $[c_1,c_2,...,c_n]$ and $[s_1,s_2,...,s_n]$. $P_1$ starts by sending the first challenge 
bit $c_1$ and $P_2$ replies with 
$c_1 \oplus s_1$. $P_1$ measures the time between sending $c_1$ and receiving the 
response. $P_1$ then replies with $c_2 \oplus s_1$. $P_2$ measures
the time between sending $c_1 \oplus s_1$ and receiving the response. This process is repeated 
$n$ times. The mutual DB procedure is considered successful if both parties verify all 
responses and match previously committed values (see \cite{sector} for more details). 
We take advantage of this optimization in constructing mutual GDB protocols.

If prover authentication is required, public key signatures can be used 
to sign challenges and responses. The verifier validates the 
signature in the last step, as shown in Figure \ref{fig:basic-db}. The protocol 
succeeds \textit{only} if the signature is valid. Public key identification schemes (e.g., Schnorr or Fiat-Shamir) can also be used 
as described in \cite{chaum-db}.

\begin{figure}[t]
\centering
\includegraphics[width=0.37\textwidth, height=0.1\textheight]{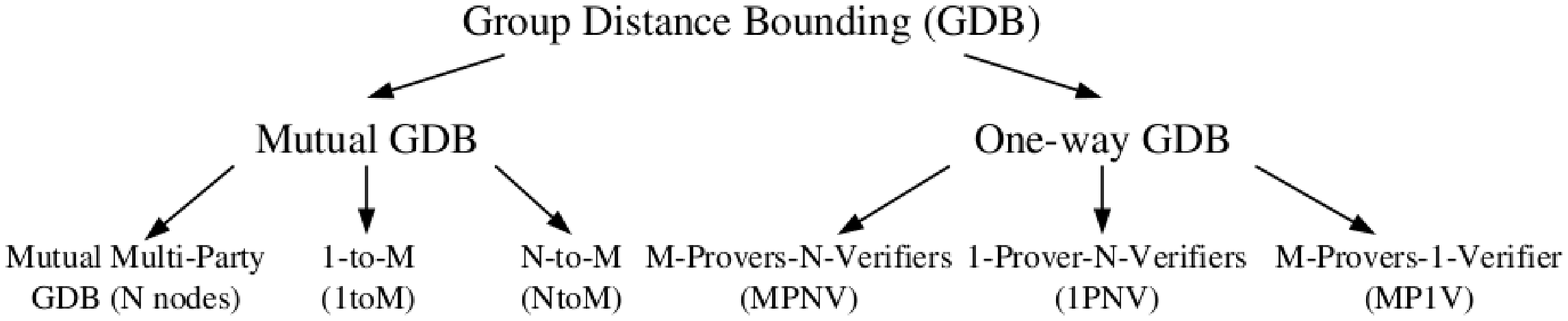}
\caption{Group Distance Bounding Variants.}
\label{fig:gdb-cases}
\end{figure}

\subsection{Problem Statement and System Model}
We first present the general GDB problem statement and its variants, then 
describe our system and adversary models.

\noindent \textbf{Problem Statement:} In general, GDB 
involves \textit{one or more provers} interacting with \textit{one or
more verifiers}. The goal of verifiers is to accurately and securely establish 
distance bounds (DBs) to provers and, optionally, authenticate them. 
Provers are generally untrusted, i.e., they may behave maliciously 
by reporting false distances and identities. Each device can be a prover, a verifier or both, i.e., 
we take into account both one-way and mutual DB. We consider three GDB cases (see
also Figure \ref{fig:gdb-cases}):\\
\textit{1- MPNV:} $N$ verifiers establish DBs to $M$ provers.\\
\textit{2- 1PNV:} $N$ verifiers establish DBs to a single prover.\\
\textit{3- MP1V:}  A single verifier establishes DBs to $M$ 
provers.

 \noindent In mutual GDB, the two special cases ($1PNV$ and $MP1V$) are equivalent and are 
called (1-to-M). In addition, there is a case where $N$ peer nodes are required to 
establish mutual DB with each other; we call it {\em mutual multi-party} GDB.

\noindent \textbf{System Model:} We make the following assumptions:

\noindent \textit{- Coverage:} All devices are within each others' transmission range. This is a 
common assumption in all DB literature, e.g., \cite{chaum-db,Kuhn10uwbdb,sec-db-loc-preneel,shmatikov-sdm,priv-db,tippenhauer09idbased}.

\noindent \textit{- Accuracy:} Each device can implement distance bounding, 
i.e., is capable of fast and accurate processing - on the order of nanoseconds\footnote{
Possible using off-the-shelf electronics as in 
\cite{kasper-db-realization-ss2010} or using UWB ranging platforms 
e.g.\cite{MSSI, Kuhn10uwbdb}.}.

\noindent \textit{- Keys:} Each device has a public/private key pair and a certificate 
 binding the public key to its identity. (Applies only if authentication is required).

\noindent \textit{- Collusion:} Colluding provers do not reveal their secret 
keys to each other. (Applies only if authentication is required).

\noindent \textit{- Interaction between Verifiers:} In one-way GDB, verifiers 
know each others' locations or distances separating them. This is not required in mutual GDB.

\noindent \textbf{Adversary Model:}
We assume that the adversary is computationally bounded and 
can not prevent nodes within its radio range from 
receiving its transmissions (i.e., not using directional antennas). In one-way GDB, the adversary 
can only compromise provers. Verifiers trust 
each other in one-way GDB. In mutual GDB, all nodes 
are treated equally with no trust assumptions. Our adversary 
model covers the following attacks in one-way GDB settings (based 
on attacks in one-way DB \cite{chaum-db}):\\
\indent \textbf{1- Distance Fraud Attack:} A dishonest prover claims to be closer than 
it really is. (Note that a prover can always claim to be further by delaying 
responses.) The goal of this attack in one-way GDB is to shorten 
the distance from the malicious prover to one or more (or even all) verifiers.
\indent \textbf{2- Mafia Fraud Attack:} A form of a man-in-the-middle (MiTM) attack. 
The adversary, who is close to the verifier, interacts with it, while posing as the prover. 
In parallel, it interacts with the prover posing as the verifier. The goal is to 
fool the verifier into believing that the adversary is the prover located 
closer to the verifier than the actual prover. 
In one-way GDB, we consider a version of this attack where the 
adversary places one or more nodes between the prover(s) and one or more verifiers. 
The adversary aims to convince verifiers that these intermediate nodes are 
real provers which are located closer to them than actual provers.

\noindent We consider the following attacks in mutual GDB settings:\\
\indent \textbf{1- Passive Distance Fraud Attack}: In mutual GDB
with one group of \textit{N} nodes, each node has to establish $N-1$ DBs. 
We assume that the adversary can compromise at most $N-2$ nodes. 
The goal of this attack is for two (or more) \textit{un-compromised} 
nodes to establish incorrect DBs to each other.
\indent \textbf{2- Node Insertion Attack}: The adversary inserts
one or more fake nodes into the group. It succeeds 
if other ``honest'' nodes in the group accept such fake nodes as legitimate 
group members and establish DBs to them. Such DB should also be 
shorter than the real distance to these fake nodes.\\

\begin{table}[t]\scriptsize
\centering
\begin{tabular}{|l|l|} \hline
%
$DB(s)$	& Distance Bound(s) \\ \hline
$P$	& Prover \\ \hline
$V \quad (V_a,V_p)$	& Verifier (subscript denotes active or passive) \\ \hline
$DB_{x,y}$	& DB established by verifier $x$ on prover $y$ \\ \hline
$t_{x,y}$	& Time of flight between nodes $x$ and $y$ \\ \hline
$d_{x,y}$	& Distance between nodes $x$ and $y$ ($d_{x,y}=d_{y,x}$) \\ \hline
$n \quad (n_a,n_p)$	& Number of DB rounds (subscript denotes active or passive) \\ \hline
$d_a$	& Fraction of verifiers performing $n_a$ active rounds \\ \hline
$H(~)$	& Cryptographically secure hash function \\ \hline
$Pr_{ch}(X)$	& Fraction of DB rounds in which node $X$ cheats \\ \hline

\end{tabular}
\caption{Notation.}
\label{tbl:notation}
\end{table}

\section{GDB Building Blocks}
\label{sec:gdb-building-blocks}
We first introduce a new building block primitive for constructing secure and efficient 
GDB protocols, \textit{one-way passive DB}. We then consider an optimization 
to decrease number of messages in GDB protocols, \textit{interleaved one-to-many 
mutual DB}. By combining the two we construct the novel \textit{mutual multi-party 
GDB} protocol. In mutual multi-party GDB each node, in a group of 
$N$ nodes, engages in a secure mutual DB protocol with its ($N-1$) peers. 
Notation used in this paper is reflected in Table \ref{tbl:notation}.

\subsection{One-Way Passive DB \label{subsec:passive-db}}
Whenever a prover and a verifier engage in a DB protocol, some information 
about their locations and mutual distance is leaked \cite{priv-db}. 
We use this observation in the presence of multiple verifiers. 
We show that it is unnecessary for \textit{every verifier} to directly interact 
with the prover ($P$) to establish a DB. If \emph{at least one} active verifier 
($V_a$) interacts with $P$, any other passive verifier ($V_p$) can deduce 
the DB between itself and $P$ by observing messages between $P$ and $V_a$. 
We assume that $V_p$ and $V_a$ trust each other, know the 
distance separating them (or each other's locations) and are both 
required to establish a DB to $P$. We address passive DB with untrusted 
verifiers in Section \ref{subsec:passive-db-untrusted-verifrs}.

\begin{figure}[t]
\centering
\includegraphics[width=0.37\textwidth, height=0.15\textheight]{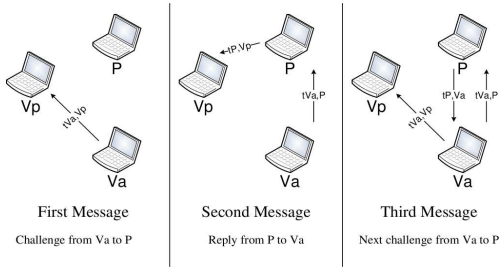}
\caption{Messages Observed by Passive Verifier.}
\label{fig:db-loc-leakage}
\end{figure}

Figure \ref{fig:db-loc-leakage} shows how $V_p$ observes timings
($T_i$) of messages exchanged in a DB protocol between $P$ and $V_a$.
$V_p$ can construct the following equations:
\begin{eqnarray}
T_1 & = & t_0+t_{V_a,V_p} \\
T_2 & = & t_0+t_{V_a,P}+\alpha_p+t_{P,V_p} \\
T_3 & = & t_0+2 \cdot t_{V_a,P}+\alpha_P+\alpha_{V_a}+t_{V_a,V_p}
\end{eqnarray} 
where $\alpha_P$ and $\alpha_{V_a}$ are processing times of $P$ and 
$V_a$, respectively (ideally $\alpha_P$ is equal to zero) and $t_0$ is the 
protocol starting time.  $V_p$ can determine time of flight for signals 
between $P$ and $V_a$ thus computing the distance between them:
\begin{equation}
d_{V_a,P}= c \cdot t_{V_a,P}= c \cdot \frac{(T_3-T_1)-\alpha_P-\alpha_{V_a}}{2}
\label{eqn:dist-time-between-active-verifier-prover}
\end{equation} 
Where $c$ denotes speed of light. For $V_a$ (and $V_p$) to measure the distance between itself and $P$, $\alpha_P$ 
must be negligible (or constant) and known\footnote{Common assumption in DB literature, e.g., 
\cite{chaum-db,Kuhn10uwbdb,sec-db-loc-preneel,shmatikov-sdm,priv-db,tippenhauer09idbased}.}. 

\textit{Overview of establishing a passive DB:} $V_p$ uses time difference of arrival (TDoA) of three messages, its own 
location and $V_a$'s location to construct the locus of $P$'s possible locations (a hyperbola similar to other TDoA techniques \cite{tdoa-positioning}). 
$V_p$ then determines the distance between $V_a$ and $P$ 
(as shown in Equation \ref{eqn:dist-time-between-active-verifier-prover}) and
constructs a circle with a radius equal to that distance. This circle 
intersects with $P$'s location locus at two points ($s_1$ and $s_2$). $V_p$ computes 
DB to $P$ as the distance between itself and $s_1$ (or $s_2$)\footnote{If $V_p$ does not know $V_a$'s exact location 
but only the distance to $V_a$, then, instead of a sector of a circle, $V_p$ 
obtains an area between two circles with radii corresponding to furthest and 
closest points to $V_p$ on the hyperbola. In that case the larger radius will be used as a DB to $P$.}.

\begin{figure*}
\centering
\subfigure[Correct Passive DB]{
\includegraphics[scale=0.3, angle =-90]{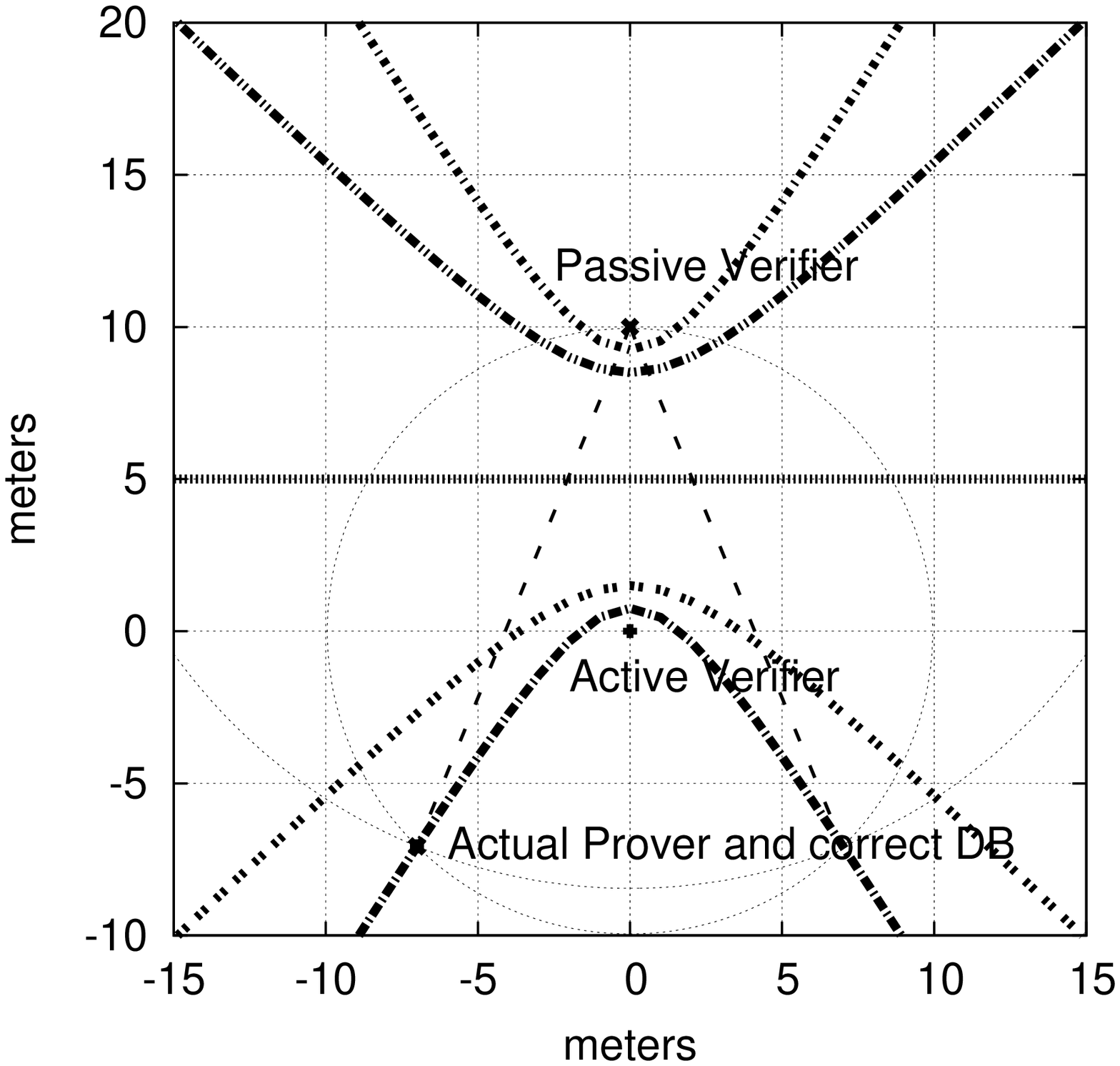}
\label{fig:correct-passive-db}
}
\subfigure[Incorrect Passive DB]{
\includegraphics[scale=0.3, angle =-90]{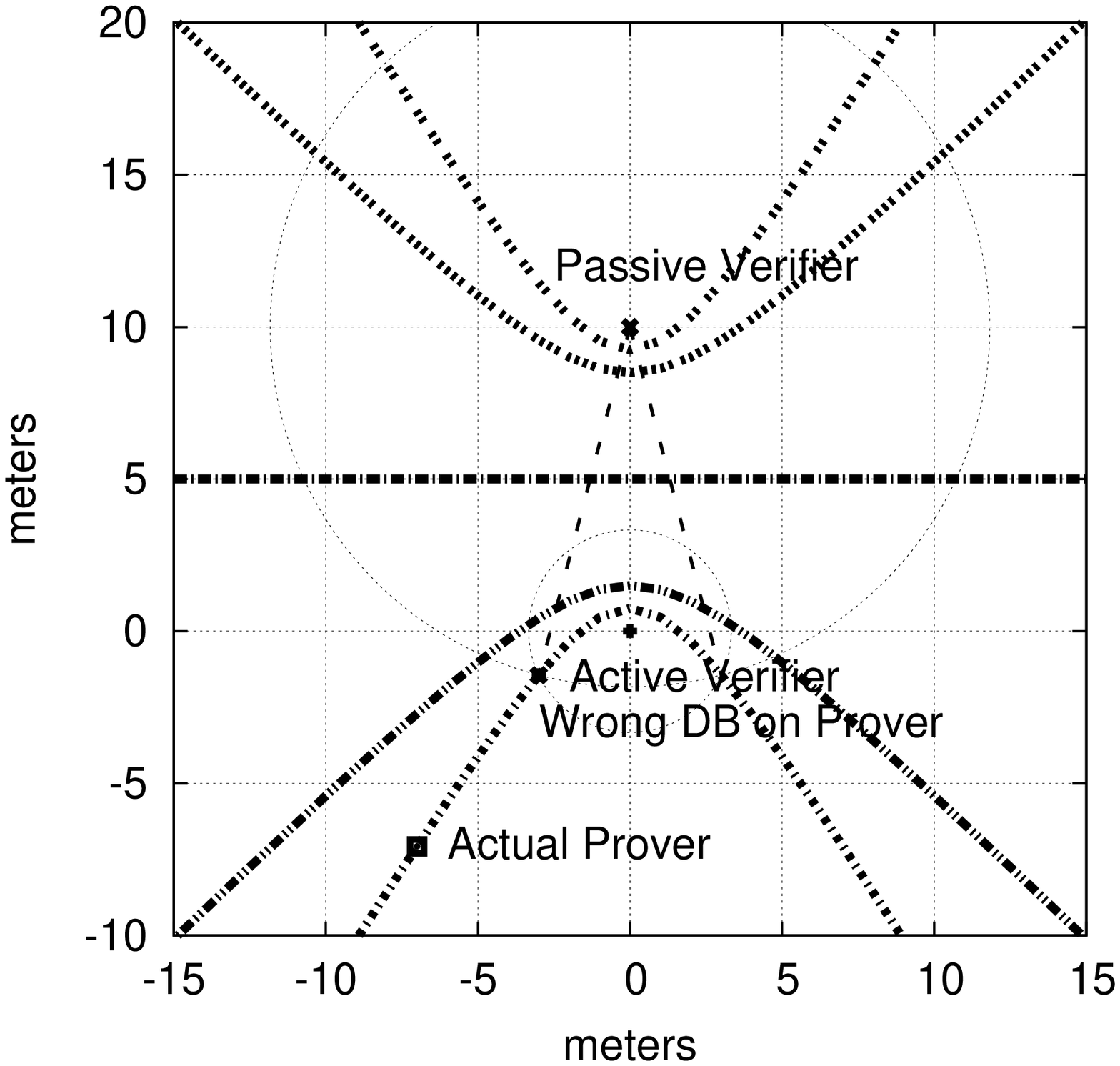}
\label{fig:wrong-passive-db}
}

\label{fig:correct-wrong-passive-db}
\caption{Establishing a Correct and Incorrect Passive DB.}
\end{figure*}

 \textit{Details of establishing a passive DB:} We now demonstrate the details 
 of the procedure. We also show that if $P$ manages 
to cheat and shorten the passive DB established by $V_p$, 
then the active DB established by $V_a$ \textit{must} also be shortened. 
Since the DB established by $V_a$ can 
not be shortened, a passive DB is as secure as the active DB established between 
$V_a$ and $P$. Suppose $V_a$ is located at $(x_a,y_a)$ and $V_p$ is at $(x_p,y_p)$. 
$V_p$ knows its own location and that of $V_a$ (hence the distance $d_{V_a,V_p}$).  Without 
loss of generality we assume $(x_a,y_a)=(0,0)$ to be the origin of a coordinate system. It follows that:
\begin{equation}
d_{V_a,V_p}=\sqrt{(x_p-x_a)^2 + (y_p-y_a)^2}= \sqrt{(x_p)^2 + (y_p)^2}
\end{equation}
We further assume that $P$ is at $(x,y)$. $V_p$ also knows that:
\begin{equation}
d_{V_p,P}=\sqrt{(x-x_p)^2 + (y-y_p)^2}=\sqrt{(x)^2 + (d_{V_a,V_p}-y)^2} 
\end{equation}
\begin{equation}
d_{V_a,P}=\sqrt{(x-x_a)^2 + (y-y_a)^2}= \sqrt{(x)^2 + (y)^2}
\end{equation}

If three messages as in Figure \ref{fig:db-loc-leakage} are received at times: $T_1,T_2$ and $T_3$, 
respectively, $V_p$ computes $d_{V_a,P}$ as shown in Equations  
\ref{eqn:dist-time-between-active-verifier-prover}. $V_p$ also computes:
\begin{equation}
c \cdot (T_2-T_1)= c \cdot \delta_1 =d_{V_a,P}+c \cdot \alpha_P+d_{P,V_p}-d_{V_a,V_p}
\end{equation}
Where $c$ is speed of light. However, since $d_{V_a,P}$ (Equations  
\ref{eqn:dist-time-between-active-verifier-prover}) and $d_{V_a,V_p}$ (verifiers know distances between them) are known, $V_p$ obtains:
\begin{eqnarray}
\Gamma=c \cdot (\delta_1-\alpha_P)+d_{V_a,V_p}=d_{V_a,P}+d_{V_p,P} = \nonumber \\
\sqrt{(x)^2 + (y)^2} + \sqrt{(x)^2 + (d_{V_a,V_p}-y)^2}
\end{eqnarray}

Which yields the following formula for the locus of $P$'s possible location (which lies on a 
hyperbola due to TDoA \cite{tdoa-positioning}):
\begin{equation}
y=\dfrac{d_{V_a,V_p} \sqrt{(d_{V_a,V_p}^2-\Gamma^2)} \pm \Gamma \cdot \sqrt{(4x^2 + d_{V_a,V_p}^2 - \Gamma^2)}}{2 \sqrt{(d_{V_a,V_p}^2 - \Gamma^2)}}
\label{eqn:hyperbola-eqn}
\end{equation}

Note that $DB_{V_a,P}=d_{V_a,P}$ 
 is an upper bound on the distance between $P$ and $V_a$. Using $d_{V_a,P}$, $V_p$ can construct another equation 
for the locus of $P$'s possible location (a circle around $V_a$ with radius $d_{V_a,P}$):
\begin{equation}
 (x-x_a)^2 + (y-y_a)^2 = (d_{V_a,P})^2
\label{eqn:va-p-circle-eqn}
\end{equation}

$V_p$ can now establish a passive DB using the intersection of both loci (i.e., solving both equations \ref{eqn:va-p-circle-eqn} and 
\ref{eqn:hyperbola-eqn}). This $DB$ is the distance between $V_p$'s own location $(x_p,y_p)$ and the 
intersection of P's loci described by equations \ref{eqn:va-p-circle-eqn} and \ref{eqn:hyperbola-eqn}. This DB ($DB_{V_p,P}=d_{V_p,P}$) will 
only be in a sector of a circle, not in the entire circle as in the case of an active DB.

Substituting $x=x_a + \sqrt{ (d_{V_a,P})^2 - (y-y_a)^2}$ (from equation \ref{eqn:va-p-circle-eqn}) into 
equation \ref{eqn:hyperbola-eqn}, the y-coordinate of 
$P$'s location becomes: $y\propto(d_{V_a,P})$ (same for $P$'s x-coordinate). For 
$V_p$ to compute a wrong (shorter) $DB$ to $P$, it has to have computed a shorter $d_{V_a,P}$. A shorter $d_{V_a,P}$ 
requires $DB_{V_a,P}$ to have been computed shorter than the actual distance between $P$ and $V_a$ 
(which is not possible as shown in Section \ref{subsec:db-overview}). 

To better illustrate this, Figures \ref{fig:correct-passive-db} 
and \ref{fig:wrong-passive-db} show an example scenario. $P$ (labeled Actual Prover in Figures) at ($-7,-7$) is on one of several possible 
hyperbolas. The DB from $V_a$ at ($0,0$) to $P$ is shown as a circle around $V_a$ (labeled as Active Verifier in Figures). 
If $P$ somehow cheats so that the passive DB is shorter, then this would require that the circle drawn 
around $V_p$ at ($0,10$) intersects the hyperbola at a point (($-3,-1.5$) in 
Figure \ref{fig:wrong-passive-db}) close to $V_p$. This point will be inside the 
circle established by $V_a$. If this is the case then the DB computed by $V_a$
has to be shorter than the actual distance to $P$ which is not possible. 
If $V_p$  actively engages in a DB protocol with $V_a$, it would get 
the circle shown around it in Figure \ref{fig:correct-passive-db}. However, in this passive case, it gets a sector of that circle, 
which is the arc connecting the two points (($-7,-7$) and ($7,-7$)) where the computed hyperbola intersects
the circle around the active verifier. We have shown in Section \ref{subsec:db-overview} how active DB prevents the distance fraud attack.
Since passive DB is as secure as active DB, it will prevent the 
distance fraud attack. Adding authentication to passive DB 
prevents the mafia-fraud attack because an attacker will not be able to authenticate itself 
to a passive verifier unless it also does to an active one. A passive verifier can 
utilize the same authentication mechanism as an active verifier.  
Active verifiers can use public key signatures (or public key 
identification schemes) to authenticate provers, as described in Section \ref{subsec:db-overview}. 
All necessary information (commitment, challenges, 
responses and signatures) required 
to authenticate provers also reach passive verifiers. The only disadvantage is 
that a passive verifier does not send its own challenges. Passive DB remains secure 
because it assumes trusted active verifiers. If that is not the case, mutual 
one-to-many DB or mutual multi-party GDB can be used.

\subsection{Interleaved One-to-Many Mutual DB}
\label{subsec:interleaved-oneto-many-mutual-DB}
When one node
engages in mutual DB with $M$ other nodes, \textit{one-to-many mutual DB}, the number 
of required messages can be reduced by interleaving challenges and responses to 
different nodes. We label the ``one'' node in this 
case the initiator ($P_i$) and the other ``many'' nodes (M) the ``participants'' ($P_j$, 
$j\in \{1,...,M\}$). $P_i$ performs 
mutual DB with each $P_j$; however, the last message of the 
interaction with one $P_j$ is used as the first challenge of the 
interaction with $P_{j+1}$. This process can 
be generalized for $M$ parties, one initiator and $n$ rounds, 
resulting in a protocol with $n \cdot (2M +1)$ messages. This would 
have required $n \cdot (4M)$ or $n \cdot (3M)$ messages if pairwise 
single prover single verifier DB or interleaved single prover single verifier DB were used respectively.

\subsection{Mutual Multi-Party GDB \label{subsec:mutualmpDB}}
The obvious approach to establish mutual DBs between every pair of nodes, in a group of $N$ nodes, 
is to perform it sequentially between each pair. This 
requires $2n \cdot N \cdot (N-1)$ messages and is insecure. A malicious node, acting as a prover, 
can selectively delay messages to another specific node acting as a verifier. This yields a 
larger DB, to that node only, and results in false localizations if multilateration is used as shown in \cite{db-wisec09}. One can 
interleave challenges and responses to reduce the number of messages to $\frac{(2n+1) \cdot N \cdot (N-1)}{2}$, but selective delaying of 
responses will still be possible. Our protocol, \emph{mutual multi-party GDB}, relies on 
the broadcast nature of the wireless channel and takes advantage of 
message overhearing and appropriate timing of challenges and responses. \textit{All} nodes simultaneously engage in 
the \textit{same protocol}. The protocol combines passive DB and interleaving of challenges and responses (similar to Section \ref{subsec:interleaved-oneto-many-mutual-DB})
 to reduce message complexity from $O(N^2)$ to $O(N)$ without sacrificing security. 

\begin{figure}[t]
\centering
\includegraphics[width=0.25\textwidth, height=0.15\textheight]{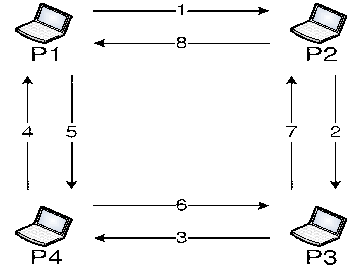}
\caption{Mutual Multi-Party GDB Example}
\label{fig:multi-party-mutual-db}
\end{figure}

We begin with a simple four-node example, shown in Figure \ref{fig:multi-party-mutual-db}. 
Each node ($k$) first generate $n$ random bit strings ($b_{i,k}$), each of length $l$. 
Each node broadcasts a commitment to these bit strings. These 
commitments are hashed and used by nodes to order themselves in a logical ring.
This ordering determines the sequence in which nodes send and respond to challenges.
In Figure \ref{fig:multi-party-mutual-db} nodes order themselves 
clock-wise starting from $P_1$ to $P_4$. $P_1$ starts and sends the first of its generated 
bits strings ($b_{1,1}$) as a challenge to its left logical neighbor $P_2$ (message 1). $P_2$ computes and sends 
the reply ($rp_{2,1}$) to $P_1$ using its own first bit string ($b_{1,2}$)
and the challenge bit string ($b_{1,1}$) received from $P_1$, i.e., message 2 is $rp_{2,1}=b_{1,1} \oplus b_{1,2}$. $P_3$ 
uses the reply from $P_2$ to $P_1$ as a challenge and computes its own reply to $P_2$ ($rp_{3,2}=b_{1,3} 
\oplus rp_{2,1}$) and broadcasts it as message 3. $P_4$ uses this 
reply ($rp_{3,2}$) as a challenge from $P_3$ and computes its own reply ($rp_{4,3}=b_{1,4} 
\oplus rp_{3,2}$) and sends it to $P_1$  as message 4. $P_1$ then replies to $P_4$ with message 5 containing $rp_{1,4}=b_{2,1} 
\oplus rp_{4,3}$. The process is repeated again counter-clock wise but using the second bit string. All 
challenges and responses are broadcast and all nodes receive and record them. Nodes only respond to challenges 
from their immediate logical neighbors. Once a node computes all required DBs, it 
broadcasts them with a hash of all received challenges and responses (and optionally signs 
them if authentication is required). This will require four additional messages. We note here that each 
node independently computes DBs to other nodes based on linear equations constructed from the reception times as illustrated in 
Linear equations can be solved with standard automated methods, e.g., Gauss elimination or Gauss-Jordan elimination.
Table \ref{tab:multi-party-mutual-db-example}. Nodes \textit{do not} rely on any reported 
measurements from other nodes, hence the same model of distrusting a node acting as a 
prover as in the original DB protocol holds.

\begin{figure}[t]
\centering
\includegraphics[width=0.4\textwidth, height=0.17\textheight]{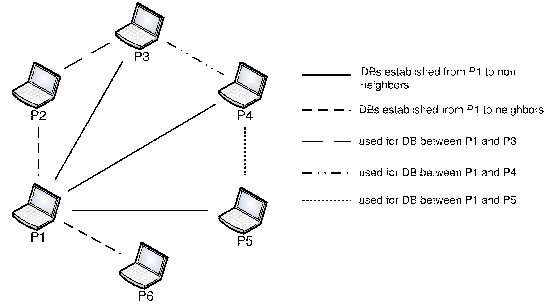}
\caption{Breaking down Mutual Multi-party GDB into Passive DBs.}
\label{fig:multi-party-mutual-db-sec-prf}
\end{figure}

Any mutual DB protocol for four nodes will require four commit 
(and four de-commit) messages which are not shown in 
Figure \ref{fig:multi-party-mutual-db}. The main difference is 
in number of messages in the rapid bit exchange phase. In Figure \ref{fig:multi-party-mutual-db} a total 
of $8$ messages are required in that phase, in the case of sequential pairwise DB 
$24$ will be needed and $18$ in case of sequential DB with interleaving. 
The process can be generalized to the case of $N$ nodes. The total number of 
messages for the general case of $N$ nodes and $n$ rounds in the rapid bit exchange phase 
is\footnote{This can be derived by analyzing the construction of linear equations from observing messages. Any node 
can construct $2N-2$ independent equations from time of arrival of $2N$ 
consecutive messages (as each node sends two messages). These equations have $2N-2$ unknowns, $N$ unknowns for time of flight 
between pairs of neighboring nodes and $N-3$ between the observing node 
and every other node (see in Figure 
\ref{fig:multi-party-mutual-db-sec-prf}). There's an additional unknown, the variable $t_0$, corresponding protocol starting time. 
These equations can be solved resulting in a unique solution.}: $n \cdot (2N)$. Additionally, $2N$ messages are required for the 
commitments and decommitments to make sure that every node has used the random bits it generated and has 
computed DBs correctly.

 
\begin{table*}[t]\scriptsize
\centering
\begin{tabular}{|l|l|l|l|l|} \hline
\textbf{Msg}	& \textbf{Participant 1 ($P_1$)}	& \textbf{Participant 2 ($P_2$)}	& \textbf{Participant 3 ($P_3$)} & \textbf{Participant 4 ($P_4$)} \\ \hline

1	& Sender & $T_1 =  t_0 + t_{P_1,P_2}$ & $T_1 =  t_0 + t_{P_1,P_3}$ & $T_1 =  t_0 + t_{P_1,P_4}$ \\ \hline

2	& $T_2 = 2 t_{P_1,P_2} \rightarrow DB_{P_1,P_2}$ & Sender & $T_2 =  t_0 + t_{P_1,P_2} + t_{P_2,P_4} + t_{P_2,P_3} $ & $T_2 =  t_0 + t_{P_1,P_2} $ \\ \hline

3	& $T_3= t_{P_1,P_2} + t_{P_2,P_3} + t_{P_3,P_1} $ & $T_3 = 2 t_{P_2,P_3} \rightarrow DB_{P_2,P_3}$  & Sender & $T_3 =  t_0 + t_{P_1,P_2} + t_{P_2,P_3} + t_{P_3,P_4} $ \\ \hline

4	& $T_4= t_{P_1,P_2} + t_{P_2,P_3} + t_{P_3,P_4} $ & $T_4 = t_{P_2,P_3} + t_{P_3,P_4} + t_{P_2,P_4} $ & $T_4 = 2 t_{P_3,P_4} \rightarrow DB_{P_3,P_4}$  & Sender \\
 & $+ t_{P_4,P_1}$ &  & & \\ \hline

5	& Sender & $T_5 = t_{P_2,P_3} + t_{P_3,P_4}$  & $T_5 = t_{P_3,P_4} $ & $T_5 = 2 t_{P_4,P_1} \rightarrow DB_{P_4,P_1}$  \\ 
	&        & $+ t_{P_4,P_1} + t_{P_1,P_2}$ & $+ t_{P_4,P_1} + t_{P_1,P_3}$ & \\ \hline

6	& $T_6 = 2 t_{P_1,P_4} \rightarrow DB_{P_1,P_4}$  & $T_6 = t_{P_2,P_3} + t_{P_3,P_4} $ &  $T_6 = 2t_{P_3,P_4} + 2t_{P_4,P_1} $ & Sender \\ 
	&       & $+ 2t_{P_4,P_1} + t_{P_4,P_2}$ &  & \\ \hline

7	& $T_7= t_{P_1,P_4} + t_{P_3,P_4} $ & $T_7 = 2t_{P_2,P_3} + 2t_{P_3,P_4} $ &  Sender & $T_7 = 2 t_{P_4,P_3} \rightarrow DB_{P_4,P_3}$ \\ 
	& $+ t_{P_3,P_1}$ & $ + 2t_{P_4,P_1} $ & &  \\ \hline
8	& $T_8= t_{P_1,P_4} + t_{P_3,P_4} $ & Sender & $T_8 = 2 t_{P_3,P_2} \rightarrow DB_{P_3,P_2}$  &  $T_8 = t_{P_4,P_3} $ \\
	& $+ t_{P_2,P_3} + t_{P_1,P_2}$& &  & $+ t_{P_3,P_2} + t_{P_2,P_4}$ \\ \hline \hline

	& (a) $T_7 + T_3 - T_4 = 2 t_{P_1,P_3} $ & (a) $T_6 - T_4 = 2 t_{P_1,P_4}$ & (a) $T_6 - 2 t_{P_3,P_4} = 2t_{P_1,P_4} $ &  (a) $T_8 + T_2 - T_3 $\\
	& $\rightarrow DB_{P_1,P_3}$ & (b) $T_7 -T_3 + T_4 - T_6 $ & (b) $T_5 - t_{P_3,P_4} $& $= 2 t_{P_4,P_2} \rightarrow DB_{P_4,P_2} $ \\
	& & $= 2 t_{P_3,P_4}$ & $- t_{P_1,P_4}  = 2 t_{P_1,P_3}$ & \\
	& & (c) $T_6 - t_{P_2,P_3} - t_{P_3,P_4}  $ & $\rightarrow DB_{P_1,P_3}$  & \\ 
	& & $- 2t_{P_1,P_4} = t_{P_2,P_4} \rightarrow DB_{P_2,P_4} $ &  & \\
	& & (d) $T_5 - t_{P_2,P_3} - t_{P_3,P_4} $ &  & \\
	& & $ - t_{P_1,P_4} = t_{P_1,P_2} \rightarrow DB_{P_2,P_1}$ &  & \\ \hline

\end{tabular}
\caption{Message Reception Times and Constructed Equations in the Mutual Multi-Party GDB Protocol for 
Figure \ref{fig:multi-party-mutual-db} ($t_{i,j}=t_{j,i}$, $t_{x,y} \rightarrow DB_{x,y}$ means that 
$DB_{x,y}$ can be directly computed from $t_{x,y}$, the last row shows additional 
computation required to establish the DBs).}

\label{tab:multi-party-mutual-db-example}
\end{table*}


\textit{Security:} The mutual multi-party GDB protocol is secure against distance-fraud and 
passive distance fraud attacks as long as the group contains at least two honest 
neighboring nodes (in the logical ring). A malicious node launching any attacks will 
be detected by these two (or more) honest neighbors because the active DB between them 
can not be influenced by \textit{any} other node. When immediate neighbors of a node exchange 
messages with their own neighbors, that node establishes passive DBs on these two-hop neighbors. These DBs are established passively and 
can not be affected by any other node because, as we have shown, passive DB is as secure as active 
DB. This process is repeated until all DBs are established.
The example in Figure \ref{fig:multi-party-mutual-db-sec-prf} shows how node $P_1$ 
uses interactions between different nodes in the group 
to establish a DB on each of them. $P_1$ establishes DB directly with its neighbors $P_2$ and 
$P_6$, it then uses the messages exchanged between $P_2$ and $P_3$ to establish a passive 
DB on $P_3$ and those between $P_3$ and $P_4$ to DB $P_4$ and between $P_4$ and $P_5$ to 
DB $P_5$. This process is carried out by each node independently during the protocol at 
different times resulting in secure DBs established to other nodes.  The 
description of Figure \ref{fig:multi-party-mutual-db-sec-prf} is simplified 
to convey the intuition. In reality, when solving linear equations constructed from TDoA 
of messages, several equations resulting from different interactions will be used in computing each DB 
to a non neighbor (details are shown in Table \ref{tab:multi-party-mutual-db-example}). For example, 
in Table \ref{tab:multi-party-mutual-db-example}, $P_1$ 
uses equations of $T_7$, $T_3$ and $T_9$ to DB $P_3$, as opposed to $T_3$ only 
as in the simplified explanation (Figure \ref{fig:multi-party-mutual-db-sec-prf}). 

To demonstrate how attacks can be detected consider how each node computes DBs 
from the arrival times of messages as shown in Table \ref{tab:multi-party-mutual-db-example}. 
Assuming $P_1$, $P_2$ and $P_4$ are honest and $P_3$ is malicious. $P_3$ can launch attacks 
by delaying its messages (number $3$ and $7$) by $\delta_1$ and $\delta_2$ respectively. 
This attack will be detected because $DB_{P_1,P_2}$ computed by $P_1$, and 
that computed by $P_2$ will not be the same. This will be detected when nodes 
broadcast their computed DBs at the end of the protocol. $P_1$ will compute 
$DB_{P_1,P_2} = T_2/2 = t_{P_1,P_2}$ (from message $2$ in the column for $P_1$), whereas $P_2$ will compute 
$DB_{P_1,P_2}=T_5 - t_{P_2,P_3} - t_{P_3,P_4} - t_{P_1,P_4} = t_{P_1,P_2} - (\delta_1 + \frac{\delta_2}{2})$ 
(from messages $5$ and $3$ and steps (a) and (b) in the last row of the column for $P_2$). 
A similar detection will occur between $P_2$ and $P_4$ but based on $DB_{P_2,P_4}$. Even if $P_3$ 
and $P_4$ are both malicious and colluding, the attack will be detected because $DB_{P_1,P_2}$ 
computed by $P_2$ will be $DB_{P_1,P_2}=t_{P_1,P_2} - \frac{\delta_2}{2}$ (assuming 
$P_3$ delays its messages by $\delta_1$ and $\delta_2$ respectively, whereas $P_4$ delays its message by 
$\delta_3$ and $\delta_4$). Variations in computed DBs can be detected 
if at least two honest nodes are neighbors in the constructed ring.

Node authentication in this protocol can be achieved using 
traditional public key signatures. Each node initially broadcasts its public key 
certificate in the commitment phase. Once all ($2nN$) protocol rounds are 
completed, each node hashes all exchanged challenges 
and responses and signs the resulting hash. Recall that all nodes receive 
all challenges and responses due to wireless broadcast. All signatures 
are then broadcasted and each node verifies $N-1$ signatures. All nodes are 
authentic if all signatures verify successfully.

\section{DB extended to Group Settings}
\label{sec:gdb-protocols}
We now show how to construct protocols for 
the two most general GDB cases: (1) $M$ provers and $N$ verifiers (MPNV) in one-way GDB, 
and (2) NtoM in mutual GDB. All other cases can be obtained by setting the values of $N$ and $M$ as desired.
For comparison, we consider  a basic GDB protocol where nodes sequentially 
engage in a na\'ive single prover single verifier (mutual) DB. In each case we propose an alternative approach 
based on passive DB, mutual multi-party GDB or one-to-many mutual DB. We assume that $n$ rounds of DB are required 
in all cases.

\subsection{One-Way MPNV GDB}
In this case nodes either act as provers or as verifiers. The goal 
at the end of the protocol is for all $N$ verifiers to have DBs to all $M$ provers. 
In a naive MPNV protocol, each prover 
interacts sequentially in $n$ rounds of DB with each verifier. This is 
repeated until \textit{all} provers have interacted with all verifiers. The 
total number of messages is: $(2n \cdot N \cdot M )$. When constructing a protocol 
for this group setting based on passive DB there are two parameters to 
consider: (1) the number of active and passive DB rounds performed by 
each verifier and (2) how the active verifiers are selected (i.e., deterministic or probabilistic). 
The second parameter does not affect the challenges and responses and how every node performs 
DB but has an effect on the security if verifiers are compromised (discussed in Section \ref{sec:perf-sec-analysis}). Active verifiers can be 
selected randomly or by any leader election protocol (e.g., \cite{manets-leader}), the rest will be passive verifiers. Other strategies 
could be explored but are out of scope of this paper.
The number of active verifiers, active rounds and how many 
perform passive rounds affects the number of messages required, the time needed for 
completing the process and the security of the DB. If all verifiers are treated equally two parameters 
can be used to describe a general protocol: (1) the number of active verifiers and (2) the number of active 
rounds by each verifier. If each of the $N$ verifiers 
is required to perform $n$ rounds of DB, we call the number of active rounds $n_a$ (and passive rounds $n_p=n-n_a$). We denote 
with $d_a$ the fraction of verifiers which perform $n_a$ active rounds. The remaining verifiers 
perform passive rounds. Each verifier will have $(d_a \cdot (N-1) \cdot n_a)$ 
opportunities to execute passive DB with each of the $M$ provers. Two interesting cases are 
obtained by setting $d_a=1/N$ and $n_a=n$, only one verifier interacts actively with all provers, 
and by setting $d_a=1$ and $n_a=n/N$ \emph{all} verifiers interact \emph{equally} with \emph{all} provers. 
By varying these two parameters ($n_a$ and $d_a$) one can obtain a protocol with the required security level and less messages 
than sequential pairwise interaction (performance and security analysis in Section \ref{sec:perf-sec-analysis}).

\subsection{Mutual NtoM GDB}
In the general case of $NtoM$ mutual DB there are two groups, 
$G_1$ and $G_2$, of $N$ and $M$ nodes respectively. \textit{All} nodes 
in $G_1$ are required to establish DBs to \textit{each of} the nodes 
in $G_2$ and vice versa, i.e., there is a total of $2 N \cdot M$ 
(but $N \cdot M$ unique) DBs to be established. There are three approaches to construct GDB 
protocols for this setting: (1) based on one-way passive DB, 
(2) mutual multi-party GDB or (3) one-to-many mutual DB.

	\textit{(1) NtoM Using Passive DB}: A fraction ($d_1$) of nodes in $G_1$, $d_1\cdot N$, will 
	establish $n_{a1}$ active and $n_{p1}$ passive DB rounds with each 
	of the $M$ nodes in $G_2$. The rest of the $(1-d_1)N$ nodes in $G_1$ establish \textit{only} 
	passive rounds. This step involves one-way DB so at the end \textit{only} nodes in $G_1$ will 
	establish DBs to nodes in $G_2$. Nodes in $G_2$ are required to perform a similar step where $d_2 \cdot M$ nodes
	establish $n_{a2}$ active and $n_{p2}$ passive rounds of DB with each 
	of the $N$ nodes in $G_1$. The rest of the $(1-d_2) M$ nodes in $G_2$ establish \textit{only} 
	passive rounds. After this step all nodes in $G_2$ will have established one-way DBs to nodes in $G_1$. This 
	protocol requires nodes in $G_1$ to trust each other and know each other's locations or distances separating them (same for $G_2$).

	\textit{(2) NtoM Using Mutual Multi-Party GDB}: All the nodes in both 
	groups can be regarded as one group of size $N+M$. The $N+M$ nodes can engage in a mutual multi-party 
	GDB protocol as shown in Section \ref{subsec:mutualmpDB}, i.e., the general case of the example of Figure \ref{fig:multi-party-mutual-db}.
	Such a protocol will require $2n (N+M)$ messages. At the end each node will have DBs to all the $N+M-1$ other nodes. Some of 
	these DBs are not required, since we assumed that nodes in $G_1$ (and $G_2$) don't perform DB on other nodes in the same group.
	
	\textit{(3) NtoM Using One-to-Many Mutual DB}: Each of the $N$ nodes in $G_1$ engages in a one-to-many mutual 
	DB protocol described in Section \ref{subsec:interleaved-oneto-many-mutual-DB} with all $M$ nodes in $G_2$. 
	This is a one-to-many mutual DB, so all nodes in $G_2$ will also establish a DB to each node 
	in $G_1$. The total number of messages in such a protocol will be: $nN \cdot (2M +1)$.

\begin{table}[t]\scriptsize
\centering
\begin{tabular}{|l|l|l|} \hline
\textbf{Setting}	& \textbf{Base Case}	& \textbf{Our Protocol}	\\
	      & \textbf{Number of Messages}	& \textbf{Number of Messages} \\ \hline

MPNV	& $2n \cdot N \cdot M$ & $(2 n_a +1)\cdot (N \cdot d_a)\cdot M$ \\ \hline

1PNV	& $(2n+1) \cdot N$ & $(2n_a+1) \cdot N\cdot d_a$  \\ \hline

MP1V	& $(2n+1) \cdot M$ & $2n + \sum_{j=1}^{M-1}(j+1)$ \\ 
& & $ \cdot (n-((M-1)-j))$ \\ \hline \hline

1toM	& $4n \cdot M$ & $n \cdot (2M +1)$  \\ \hline
NtoM	& $4 n \cdot N \cdot M$ & $2n \cdot (N+M)$ \\ \hline

\end{tabular}
\caption[Number of Messages in One-Way and Mutual GDB]{Number of Messages in GDB Protocols.}

\label{tab:num-msgs-one-way-mutual-db-all-settings}
\end{table}

\section{Performance and Security Analysis}
\label{sec:perf-sec-analysis}
We first analyze performance of proposed GDB protocols. We then consider security of active
DB in group settings, passive DB with untrusted verifiers and their combination. Our 
GDB protocols either use mutual multi-party GDB or a combination of passive and active DB. Their security 
can be understood by analyzing the underlying mechanisms and combinations thereof. 
Correctness and security of passive DB and mutual multi-party GDB are analyzed 
in Sections \ref{subsec:passive-db} and \ref{subsec:mutualmpDB} respectively.

\subsection{Performance of GDB Protocols}
Table \ref{tab:num-msgs-one-way-mutual-db-all-settings} compares number of messages 
required in one-way and mutual GDB protocols to the base case 
(running pairwise DB between nodes). Table \ref{tab:time-one-way-mutual-db-all-settings} 
shows total time required to compute all DBs. We compare against this base case because there are 
no previous proposals for GDB. Our $NtoM$ protocol in both tables is based on mutual multi-party GDB.
Our proposals require fewer messages and depend on the fraction of 
active verifiers and active rounds performed. In the MPNV case, only 
($n_a \cdot d_a$) messages are required, where $n_a$ is the fraction of active rounds performed 
by the fraction of active verifiers, $d_a$. Figure \ref{fig:num_msgs_na_da_n10_N10_M10} 
shows how the number of messages increases as a function of the fraction 
of active rounds and active verifiers (M=10 provers and N=10 verifiers and n=10 DB rounds). 
Figure \ref{fig:num_msgs_na_da_n10_N_M} shows how the number of messages varies 
with the number of provers and verifiers (to illustrate this dependency we assume that number of provers is 
the same as number of verifiers, $N=M$, and that fraction of active rounds is 
equal to fraction of active verifiers, $n_a=d_a$). In the case of 60 nodes (30 provers and 30 verifiers) 
if the fraction of active verifiers and active rounds is reduced to $0.8$, $33\%$ of messages can be saved. 
Decreasing this fraction to $0.6$ saves more than $55\%$ of messages. Similar savings are also attainable for 
lower and larger numbers of provers/verifiers.

\subsection{Security of Active DB vs Mutual Multi-Party GDB}
The probability of a single prover successfully cheating a single verifier decreases exponentially with the 
number of DB rounds ($n$) in an active DB protocol. 
For $n$ rounds a prover has $2^{-n}$ chance to successfully guess all challenge 
bits and send responses ahead of time. This tricks the verifier into measuring a 
shorter round trip time of flight\footnote{$2^{-n}$ is the 
probability for Brands-Chaum protocol \cite{chaum-db}, whereas in 
some other protocols like Hancke-Kuhn \cite{hancke-RFID-db-securecomm05}, this 
probability is $(3/4)^{-n}$.}. A Verifier in an active DB protocol does not 
have to trust any other entity. In group settings where each pair of provers 
verifiers engage in an active DB protocol, these security guarantees 
still hold. However, active DB in group settings is insecure if used for 
localization. When a prover actively interacts with each 
verifier separately, it can selectively enlarge its distance by 
delaying messages. Verifiers will incorrectly localize the prover 
using such DBs. Secure-localization schemes must \textit{always} 
require \textit{at least three} verifiers to interact 
with the prover \textit{simultaneously}. The mutual multi-party GDB protocol 
achieves this by design. In mutual multi-party GDB \textit{all} nodes participate 
simultaneously in the \textit{same} protocol and 
overhear each other's messages. A node can not selectively delay 
messages to other nodes, it can either delay them to \textit{all} nodes or \textit{none}.

\begin{table}[t]\scriptsize
\centering
\begin{tabular}{|l|l|l|} \hline
\textbf{Setting}	& \textbf{Base Case Time} & \textbf{Our Protocol Time} \\ \hline

MPNV	& $2n \cdot \sum_{i=1}^{N} \sum_{j=1}^{M} t_{V_i,P_j}$ & $(2 n_a +1) \cdot \sum_{j=1}^{d_a \cdot N} \sum_{k=1}^{M} t_{P_k,V_j}$	\\ \hline

1PNV	& $2n \cdot \sum^{N}_{i=1} t_{P,V_i}$ & $(2 n_a +1) \cdot \sum_{j=1}^{d_a \cdot N} t_{P,V_j}$ \\ \hline

MP1V	& $2n \cdot \sum_{i=1}^j{t_{V,P_i}}$ , $j\in{1,M}$ & $(n \cdot max(t_{V,P_i}))+\sum_{i=1}^{M-1}t_{V,P_i}$	\\ \hline \hline

1toM	& $4n \cdot \sum_{j=1}^M t_{P_i,P_j}$ & $2n \cdot \sum_{j=1}^{M+1} t_{P_j,P_{(j+1)mod(M+1)}}$ 	\\ \hline
NtoM	& $4n \sum_{i=1}^N \sum_{j=1}^M t_{P_i,P_j}$ & $2n \cdot (N+M) max(t_{P_i,P_j})$, \\ 
	& & $\forall i,j \in\{1,M\}$ \\ \hline

\end{tabular}
\caption[Time Required in One-Way and Mutual GDB]{Time Required in GDB Protocols.}

\label{tab:time-one-way-mutual-db-all-settings}
\end{table}

\subsection{Security of Passive DB with Untrusted Active Verifiers \label{subsec:passive-db-untrusted-verifrs}}
Passive DB with untrusted verifiers is mainly 
useful in MANETs, where nodes continuously encounter new peers. 
Passive DB is secure if the active verifier behaves honestly and is trusted 
as shown in Section \ref{subsec:passive-db}. This will be the case in a fixed 
(or mobile) verification infrastructure with prior security association, or 
under the control of one administrative entity. A malicious active verifier 
can undermine security of passive DB as follows: 

\emph{(1) Reporting a Fake Location (or Distance):} A passive 
verifier requires the exact location or the distance to the 
active one in order to be able to construct the DB as shown in 
Section \ref{subsec:passive-db}. The passive verifier will 
wrongfully compute the hyperbola (in Equation \ref{eqn:hyperbola-eqn}), 
if the active verifier reports an incorrect location or distance, leading 
to a wrong passive DB.\\
 \indent \emph{(2) Sending Early Challenges}: Even if the active 
verifier reports its location or distance correctly, it can 
send new challenges prematurely. This leads the passive verifier 
to believe that the prover is closer than it actually is 
(as shown in Figure \ref{fig:db-loc-leakage}). $V_p$ wrongfully 
computes the distance $d_{V_a,P}$ intersecting incorrectly 
with the hyperbola (as shown in Figure \ref{fig:correct-passive-db}), 
and leading to a wrong DB.\\

An essential aspect of passive DB is implicit trust that the active 
verifier is behaving honestly, i.e., not cheating by performing 
either of the previous two attacks. We devise a metric, the 
\emph{DB Correctness} $(DBC)$, to illustrate the effectiveness of passive DB 
in the presence of such attacks. We define $DBC$ for $n$ rounds of 
passive DB as follows:
\begin{equation}
\label{eqn:DBC}
DBC = 1 - 2^{n \cdot (Pr_{ch}(V_a) -1)}
\end{equation}

where $Pr_{ch}(V_a)$ is the fraction of rounds in which $V_a$ cheats. Note 
that cheating in passive DB is different than in active DB. 
In active DB, $P$ is the one cheating, whereas in passive DB we are concerned with the case where 
$V_a$ is the one cheating. When $V_a$ does not cheat in any DB rounds, $Pr_{ch}(V_a)$ will be $0$ and 
$DBC=1-2^{-n}$. When $V_a$ cheats in all DB rounds $Pr_{ch}(V_a)$ will be $1$ and $DBC=0$. 
The average correctness ($DBC_{avg}$) in a DB to a given prover, obtained by a passive 
verifier, in the case of $N$ active verifiers ($V_{a_1},V_{a_2} ... V_{a_N},$) (each 
engaging in $n_1,n_2 ... n_N$ rounds of DB respectively) is computed as the 
average of individual $DBC$ for each verifier:
\begin{equation}
\label{eqn:DBCAVG}
DBC_{avg}= \frac{N - \sum_{i=1}^{N} 2^{n_i \cdot (Pr_{ch}(V_a(i)) -1)}}{N}
\end{equation}

Figure \ref{fig:dbc_avg_passive_oneway_db} shows how $DBC_{avg}$ is 
affected by varying fraction of rounds in which active verifiers cheat (x-axis) 
and the fraction of cheating active verifiers. In the case considered 
(10 verifiers and 10 DB rounds) even if 50\% of the active verifiers cheat in 50\% 
of their rounds, the DB will be established correctly over 98\% of the time. 
As long as less than half of the active verifiers cheat in less than 90\% of their 
rounds, the DB will be correct more than 70\% of the time.

\subsection{Combined Passive/Active DB Security}
When a verifier performs $n_a$ active rounds and $n_p$ passive rounds both can be 
combined to obtain a more stable DB. We estimate the correctness in such a combined 
DB using a metric ($DBC_{a/p}$) as follows (note that both passive and 
active rounds have to result in the same DB): 

\begin{equation}
\label{eqn:DBCAP}
DBC_{a/p}= 1- ( 2^{-n_a} \cdot  \frac {\sum_{i=1}^{N} 2^{n_{p(i)} \cdot (Pr_{ch}(V_a(i)) -1)} }{N} )
\end{equation}

\begin{figure*}[t]
\centering
\subfigure[]{
\includegraphics[scale=0.16, angle=-90]{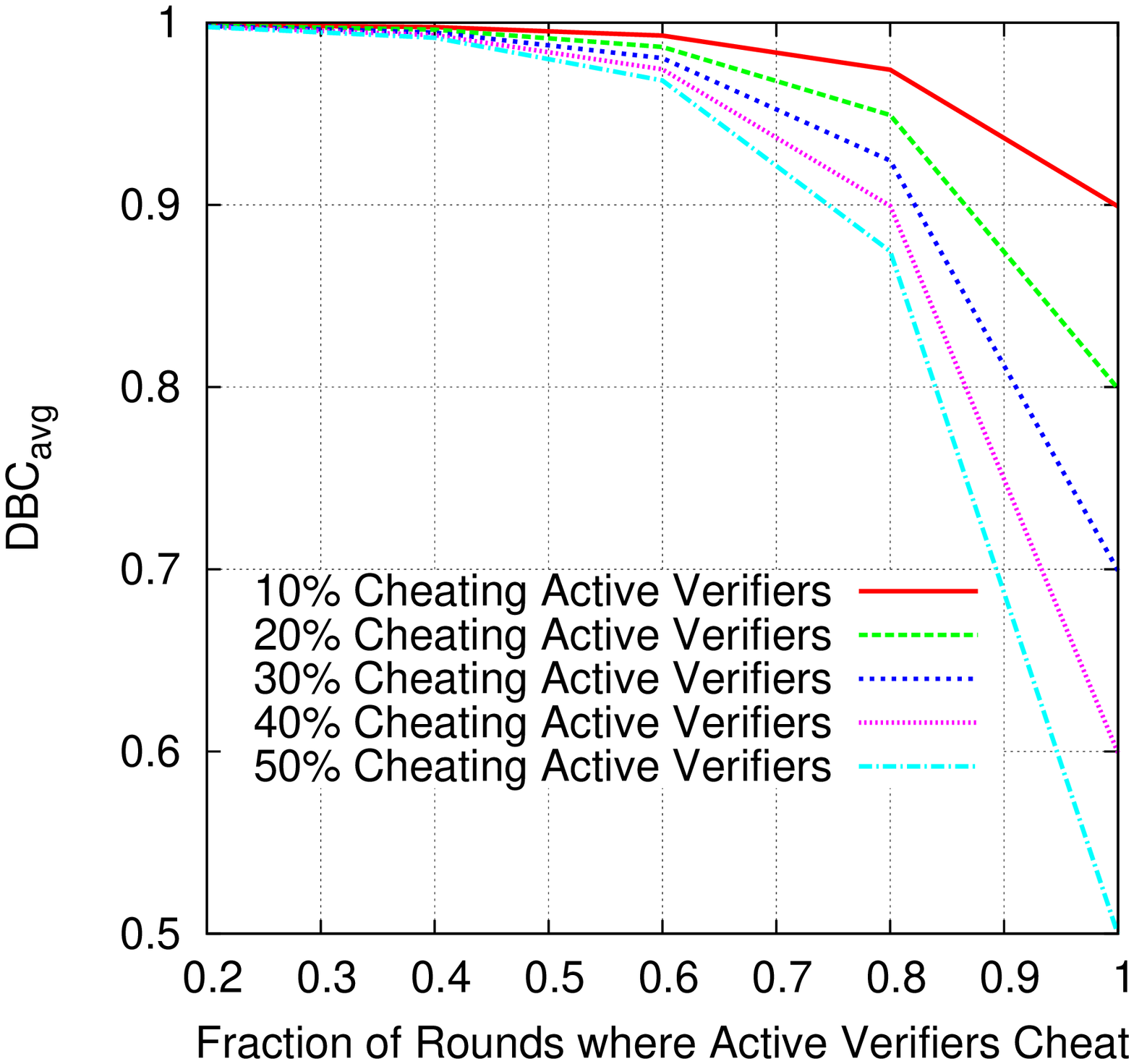}
\label{fig:dbc_avg_passive_oneway_db}
}
\centering
\subfigure[]{
\includegraphics[scale=0.16, angle=-90]{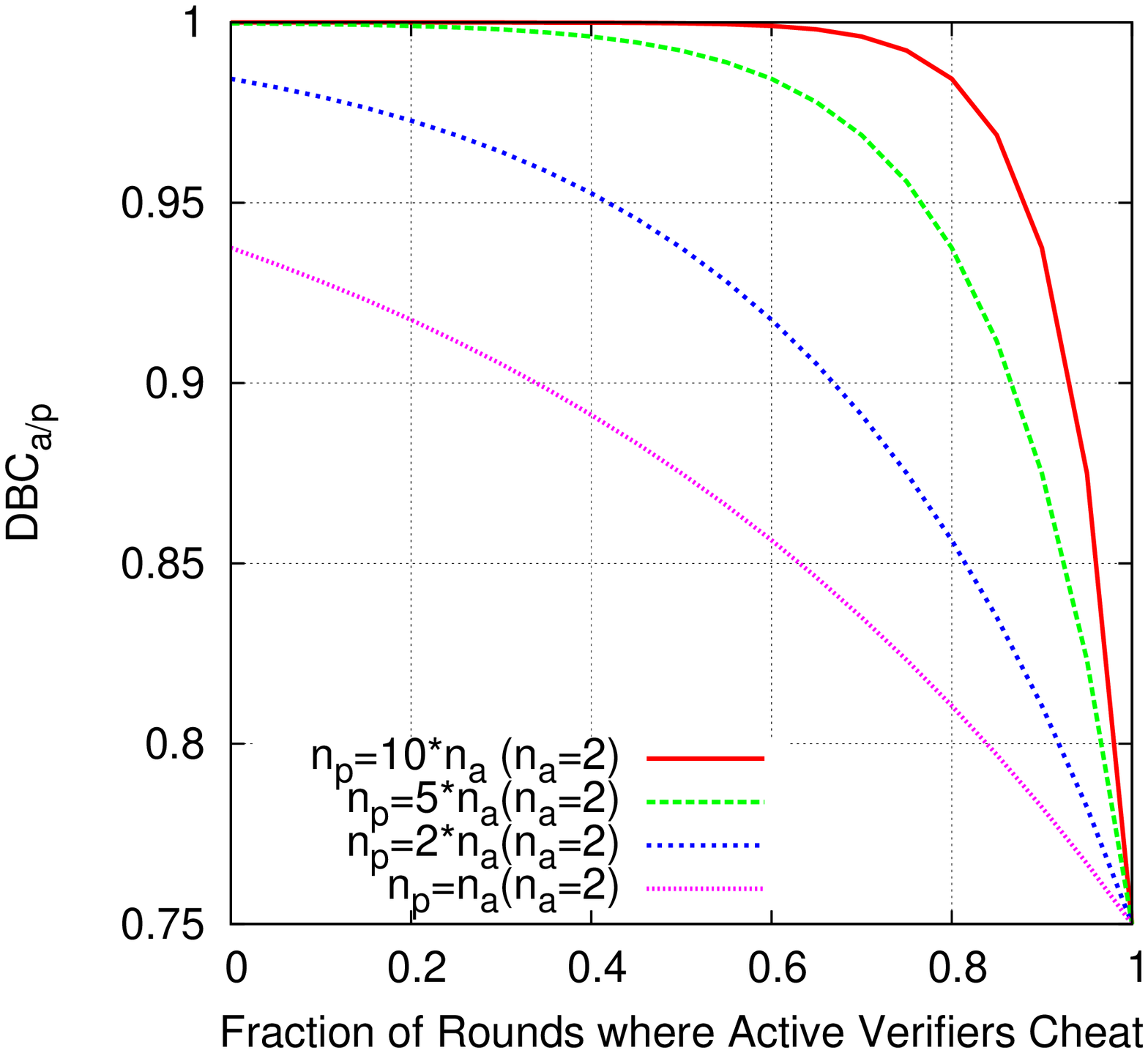}
\label{fig:dbc_active_passive_oneway_db}
}
\centering
\subfigure[]{
\includegraphics[scale=0.16, angle=-90]{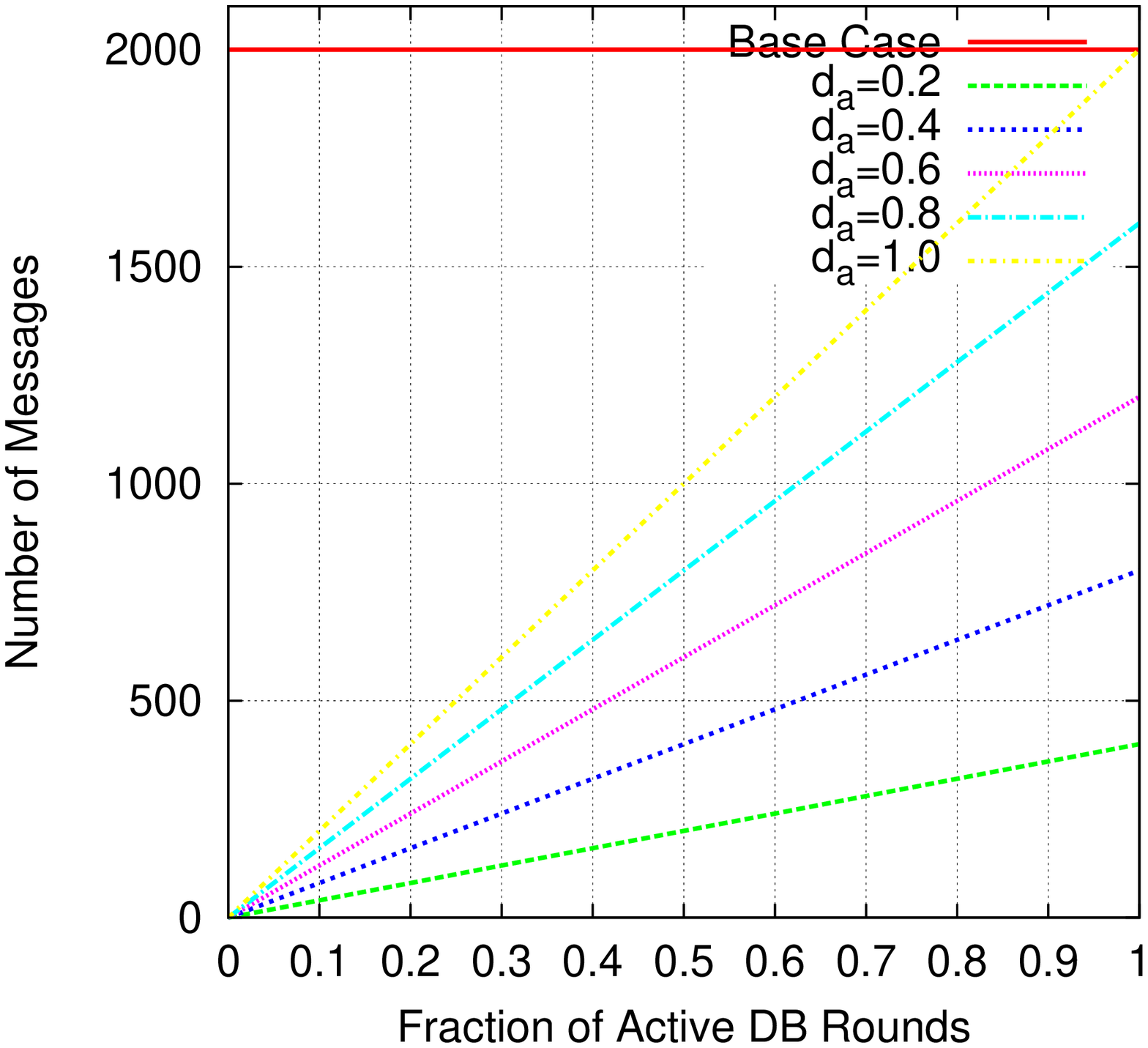}
\label{fig:num_msgs_na_da_n10_N10_M10}
}
\centering
\subfigure[]{
\includegraphics[scale=0.16, angle=-90]{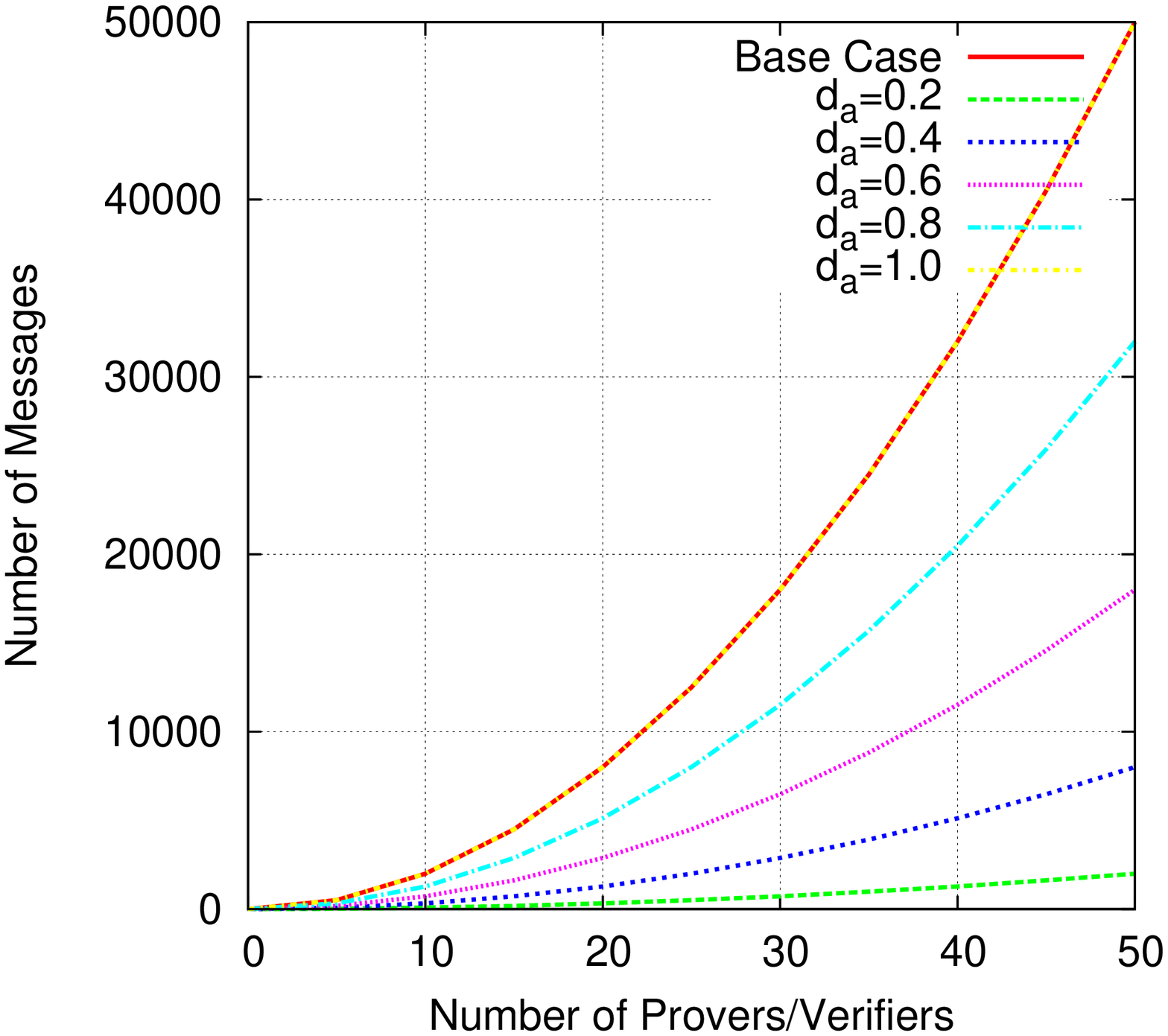}
\label{fig:num_msgs_na_da_n10_N_M}
}

\caption{(a) DBC$_{avg}$ (Equation \ref{eqn:DBCAVG}) and (b) DBC$_{a/p}$ (Equation \ref{eqn:DBCAP}) vs Probability 
of Cheating in Rounds with Ten Verifiers (N=10); (c) and (d) Number of Messages in Passive DB based MPNV Protocol\label{fig:num_msgs-mpnv}}
\end{figure*}

If all other active verifiers cheat in all their DB rounds, $DBC_{a/p}$ becomes that 
of the active rounds performed by a verifier only, i.e., $1- ( 2^{-n_a})$. Otherwise the 
likelihood of correctness of the established DB increases with any additional passive 
rounds. Figure \ref{fig:dbc_active_passive_oneway_db} shows how the $DB$ is 
affected by cheating of active verifiers during passive DB by showing how $DBC_{a/p}$ changes. In the 
case considered (10 verifiers) even if only two rounds of active ($n_a$) DB are performed and as long as 
the fraction of rounds being cheated in is less than 1 correctness of the DB captured 
by $DBC_{a/p}$ increases. Even if the probability of cheating in passive DB 
rounds is as high as 0.5, $DBC_{a/p}$ will increase to over 0.95 if 
there are four or more opportunities to do passive DB.

\section{Related Work}
\label{sec:related-work}

DB was first proposed in \cite{chaum-db} to enable 
a \textit{single verifier} to determine an upper-bound on the physical 
distance to a \textit{single prover} and authenticate it 
as summarized in Section \ref{subsec:db-overview}. Several optimizations and studies of DB were 
then considered. In particular, \cite{priv-db} studied information leakage in 
DB protocols as a privacy problem that should be avoided. In our work, we start from 
this observation to construct passive DB and the mutual multi-party GDB protocol. 
\cite{sector} proposed a mutual DB protocol by interleaving challenges and 
responses but also between a single prover and a single verifier. \cite{sec-db-loc-preneel}, 
\cite{shmatikov-sdm} and \cite{capkun-infocom05} investigated using 
DB protocols for location verification and secure localization with three verifiers. 
The setting in \cite{sec-db-loc-preneel} is a special case of 
MPNV with $M=1$ and $N=3$. \cite{wagner-sec-verif-loc} investigated the so-called 
``in-region verification" and claimed that, for certain applications, such 
as sensor networks and location-based 
access control,  in-region verification is a better match than location 
determination. \cite{db-wisec09} and \cite{pos_crypto09} considered 
collusion attacks on DB location verification protocols. 
Other work, such as \cite{uwb-tdoa-loc} looked at using time difference of arrival (TDoA) to 
determine location of transmitters. \cite{uwb-tdoa-loc} proposed 
using TDoA in the context of Ultra-Wideband (UWB). The work in 
\cite{tippenhauer09idbased,Kuhn10uwbdb} recently implemented the 
first RF based TDoA secure localization system using commercial 
off-the-shelf UWB ranging devices. DB was also studied 
in the context of ad-hoc networks (e.g., \cite{sector}), sensor 
network (e.g., \cite{meadows-db-sensors} \cite{capkun-infocom05}) and 
RFID (e.g., \cite{RFID-db-usenix-sec-07} \cite{hancke-RFID-db-securecomm05}) 
applications. Finally, DB has been used to develop secure proximity based 
access control protocols for implementable medical 
devices in \cite{rasmussenCCS09} and implemented 
using commercial off-the-shelf electronic components in \cite{kasper-db-realization-ss2010}.

To summarize, our work differs from prior results, since: (1) we introduce 
for the first time \textit{passive DB} and the \textit{mutual multi-party 
GDB protocol} which are more suitable for group 
settings, (2) we consider general GDB cases with multiple provers and multiple verifiers 
(in the one-way and mutual DB settings), (3) we study a large spectrum of possible 
protocol designs and (4) consider node authentication 
in both one-way and mutual GDB.

\section{Discussion and Conclusion}
\label{sec:disc-future-work}
This paper presents the first investigation of group distance bounding (GDB). 
GDB is a fundamental mechanism for secure operation in wireless networks where 
verifying distances between, or locations of, groups of nodes is required. 
We have shown how to construct protocols that are more efficient and secure than applying existing DB 
techniques in group settings. We made minimal assumptions about GDB settings to make our proposals 
as general as possible. However we acknowledge two open 
issues: (1) It remains an open question whether a passive verifier 
can passively establish a DB without knowing the location of (or distance to) an active 
one, while perhaps knowing other information about distances to other nodes. 
(2) We have not addressed denial-of-service attacks 
in group settings (i.e., noisy environments). We note though that single prover single verifier DB in 
noisy environments has been addressed in both the one-way and mutual cases 
in \cite{hancke-RFID-db-securecomm05,Singlee}.

\scriptsize
\bibliographystyle{plain}
\bibliography{gdb}

\begin{thebibliography}{10}

\bibitem{MSSI}
{Multispectral Solutions Inc., Urban Positioning System (UPS).}
\newblock \url{http://www.multispectral.com}.

\bibitem{mil-manet-req}
{RFC1677-Tactical Radio Frequency Communication Requirements for IPng}.
\newblock \url{http://www.faqs.org/rfcs/rfc1677.html}.

\bibitem{francillion10}
A.~Francillion {B. Danev} and S.~\v{C}apkun.
\newblock Relay attacks on passive keyless entry and start systems in modern
  cars.
\newblock In {\em Cryptology ePrint Archive: Report 2010/332}, 2010.

\bibitem{chaum-db}
S.~Brands and D.~Chaum.
\newblock Distance-bounding protocols.
\newblock In {\em EUROCRYPT '93}, pages 344--359. Springer-Verlag New York,
  Inc., 1994.

\bibitem{HAN}
E.~Callaway and P.~Gorday et.~al {et. al}.
\newblock Home networking with ieee 802.15.4: a developing standard for
  low-rate wireless personal area networks.
\newblock {\em Communications Magazine, IEEE}, 40(8):70--77, aug 2002.

\bibitem{capkun-infocom05}
S.~Capkun and J.~Hubaux.
\newblock Secure positioning of wireless devices with application to sensor
  networks.
\newblock In {\em {IEEE} {INFOCOM}}, 2005.

\bibitem{pos_crypto09}
N.~Chandran, V.~Goyal, R.~Moriarty, and R.~Ostrovsky.
\newblock Position based cryptography.
\newblock In {\em CRYPTO '09}, pages 391--407, Berlin, Heidelberg, 2009.
  Springer-Verlag.

\bibitem{db-wisec09}
Jerry~T. Chiang, Jason~J. Haas, and Yih-Chun Hu.
\newblock Secure and precise location verification using distance bounding and
  simultaneous multilateration.
\newblock In {\em ACM WiSec '09}, pages 181--192.

\bibitem{RFID-db-usenix-sec-07}
S.~Drimer and S.~Murdoch.
\newblock Keep your enemies close: distance bounding against smartcard relay
  attacks.
\newblock In {\em SS'07: Proceedings of 16th USENIX Security Symposium},
  Berkeley, CA, USA, 2007. USENIX Association.

\bibitem{oakland-automotive-sec-2010}
F.~Roesner et~al. {A. Czeskis}.
\newblock Experimental security analysis of a modern automobile.
\newblock {\em In IEEE Symposium on Security and Privacy}, 0:447--462, 2010.

\bibitem{gangs}
O.~Chen et~al. {C. Chen}.
\newblock Gangs: gather, authenticate 'n group securely.
\newblock In {\em MobiCom'08}, pages 92--103, New York, NY, USA. ACM.

\bibitem{tdoa-positioning}
Fredrik Gunnarsson.
\newblock Positioning using time-difference of arrival measurements.
\newblock In {\em In Proceedings of the IEEE International Conference on
  Acoustics, Speech, and Signal Processing}, 2003.

\bibitem{hancke-RFID-db-securecomm05}
G.~Hancke and M.~Kuhn.
\newblock An rfid distance bounding protocol.
\newblock In {\em SECURECOMM '05}, pages 67--73, Washington, DC, USA, 2005.
  IEEE Computer Society.

\bibitem{Kuhn10uwbdb}
H.~Luecken {M. Kuhn} and N.~Tippenhauer.
\newblock {UWB} impulse radio based distance bounding.
\newblock In {\em Proceedings of the Workshop on Positioning, Navigation and
  Communication (WPNC)}, 2010.

\bibitem{meadows-db-sensors}
C.~Meadows, P.~Syverson, and L.~Chang.
\newblock Towards more efficient distance bounding protocols for use in sensor
  networks.
\newblock In {\em Securecomm and Workshops, 2006}, pages 1--5, 28 2006-Sept. 1
  2006.

\bibitem{manets-leader}
J.~Welch {N. Malpani} and N.~Vaidya.
\newblock Leader election algorithms for mobile ad hoc networks.
\newblock In {\em DIALM'00}, pages 96--103.

\bibitem{rasmussenCCS09}
K.~Rasmussen, C.~Castelluccia, T.~Heydt-Benjamin, and S.~\v{C}apkun.
\newblock Proximity-based access control for implantable medical devices.
\newblock In {\em ACM CCS'09}, 2009.

\bibitem{priv-db}
K.~Rasmussen and S.~\v{C}apkun.
\newblock Location privacy of distance bounding protocols.
\newblock In {\em ACM CCS '08}, pages 149--160, 2008.

\bibitem{kasper-db-realization-ss2010}
K.~Rasmussen and S.~\v{C}apkun.
\newblock Realization of rf distance bounding.
\newblock In {\em Proceedings of the USENIX Security Symposium}, 2010.

\bibitem{wagner-sec-verif-loc}
N.~Sastry, U.~Shankar, and D.~Wagner.
\newblock Secure verification of location claims.
\newblock In {\em WiSe '03: Proceedings of the 2nd ACM workshop on Wireless
  security}, New York, NY, USA, 2003. ACM.

\bibitem{shmatikov-sdm}
V.~Shmatikov and M.~Wang.
\newblock Secure verification of location claims with simultaneous distance
  modification.
\newblock In {\em ASIAN}, 2007.

\bibitem{Singlee}
D.~Singelee and B.~Preneel.
\newblock Distance bounding in noisy environments.
\newblock In {\em ESAS 2007}, volume 4572 of {\em Lecture Notes in Computer
  Science}, pages 101--115, Cambridge,UK. Springer-Verlag.

\bibitem{sec-db-loc-preneel}
D.~Singelee and B.~Preneel.
\newblock Location verification using secure distance bounding protocols.
\newblock In {\em Mobile Adhoc and Sensor Systems Conference, 2005. IEEE
  International Conference on}, Nov. 2005.

\bibitem{tippenhauer09idbased}
N.~Tippenhauer and S.~\v{C}apkun.
\newblock Id-based secure distance bounding and localization.
\newblock In {\em In Proceedings of ESORICS (European Symposium on Research in
  Computer Security)}, 2009.

\bibitem{sector}
S.~\v{C}apkun. L.~Butty{\'a}n and J.~Hubaux.
\newblock Sector: secure tracking of node encounters in multi-hop wireless
  networks.
\newblock In {\em ACM SASN'03}, pages 21--32, New York, NY, USA. ACM.

\bibitem{uwb-tdoa-loc}
D.~Young, C.~Keller, D.~Bliss, and K.~Forsythe.
\newblock Ultra-wideband (uwb) transmitter location using time difference of
  arrival (tdoa) techniques.
\newblock volume~2, pages 1225--1229 Vol.2, Nov. 2003.

\end{thebibliography}

\end{document}